\documentclass[english,aps,preprint]{revtex4}
\usepackage[T1]{fontenc}
\usepackage[latin9]{inputenc}
\usepackage{float}
\usepackage{bm}
\usepackage{amsmath}
\usepackage{amssymb}
\usepackage{graphicx}
\usepackage{esint}
\usepackage {ushort}
\usepackage{xcolor}
\usepackage{ulem}

\makeatletter

\@ifundefined{textcolor}{}
{%
 \definecolor{BLACK}{gray}{0}
 \definecolor{WHITE}{gray}{1}
 \definecolor{RED}{rgb}{1,0,0}
 \definecolor{GREEN}{rgb}{0,1,0}
 \definecolor{BLUE}{rgb}{0,0,1}
 \definecolor{CYAN}{cmyk}{1,0,0,0}
 \definecolor{MAGENTA}{cmyk}{0,1,0,0}
 \definecolor{YELLOW}{cmyk}{0,0,1,0}
}

\begin{document}


\vspace*{1cm}

\title{Role of the action function in defining electronic wavefunctions for large systems}

\author{P. Fulde}
\email{fulde@pks.mpg.de}
\affiliation{Max-Planck-Institut f\"ur Physik komplexer Systeme, N\"othnitzer Stra\ss e 38, 01187 Dresden, Germany}
\date{\today}

\vspace*{2cm}

\begin{abstract}
The dimension of the Hilbert space needed for the description of an interacting electron system increases exponentially with electron number $N$. As pointed out by W. Kohn this exponential wall problem (EWP) limits the concept of many-electron wavefunctions based on solutions of Schr\"odinger's equation to $N \leq N_0$, where $N_0 \approx 10^3$ \cite{Kohn1999}. This limitation does not hold when the electronic interactions are neglected or treated in a mean-field approximation. The EWP has directed electronic structure calculations for solids to methods like density-functional theory which avoid dealing with many-electron wavefunctions. We show that the highly unsatisfactory limitation of many-electron wavefunctions to $N \leq N_0$ can be overcome by generalizing their definition. We show for the ground state of a large system that it is the logarithm of the solution of Schr\"odinger's equation which should be used to characterize this system. In this case the wavefunctions of independent subsystems add up rather than multiply. This feature is realized by the action function. It provides a simple physical picture for the resolution of the EWP. 
\end{abstract}

\maketitle

An electron has a particle as well as a wave character. This dual feature is well described by Schr\"odinger's equation \cite{Schroedinger1926}. When dealing with $N$ interacting electrons the usual assumption is that the same holds true, except that solving the equation by approximations becomes increasingly more difficult. However, this is not true, though. The reason is that the dimension of Hilbert space spanned by the functions which are needed for the description of the system increases exponentially with electron number $N$. This exponential increase limits the concept of Schr\"odinger's wavefunction. Indeed, the following statement was made by Walter Kohn in his Nobel lecture \cite{Kohn1999}: for a system of $N > N_0$ interacting electrons, where typically $N_0 \approx 10^3$ is a threshold value 
a wavefunction $\psi \left( {\bf r}_1 \sigma_1, {\bf r}_2 \sigma_2,  \dots, {\bf r}_N \sigma_N \right)$ is no longer a legitimate scientific concept. The reason is the following:

Clearly Schr\"odingers equation for a many-electron problem cannot be solved exactly. Therefore one must be able to replace e.g., the unknown exact ground-state wavefunction in a controlled way by an approximate one. This is not possible though when $N \geq N_0$. Consider the ground state $| \psi_0 \rangle$ of a many-electron system. It is noticed that the overlap with any approximate state $| \psi_0^{\rm app} \rangle$ vanishes exponentially for large $N$, i.e.,
\begin{equation}
\mid \left< \psi_0 | \psi_0^{\rm app}\right> \mid \leq \left( 1 - \epsilon \right)^N ~~~.
\label{eq:01}
\end{equation}
Here $\epsilon$ is an acceptable inaccuracy in the description of an electron in the presence of the interactions with the other electrons. This implies that any $| \psi_0^{\rm app} \rangle$ is orthogonal to $| \psi_0 \rangle$ for all purposes when $N \geq N_0$. This finite threshold depends on the computer power and on the used correlation method but a value of $N_0 \approx 10^3$ seems appropriate. This does not prevent us from inventing different trial wavfunctions $| \psi_{\rm trial} \rangle$ and asking which of them has an energy expectation value as low as possible. Yet, these trial functions are unrelated to $| \psi_0 \rangle$ since they are all orthogonal to the latter, i.e., $\langle \psi_0 | \psi_{\rm trial} \rangle = 0$ when $N > N_0$.

Another requirement for a wavefunction is that it can be documented. Here $| \psi_0 \rangle$ would need to specify an exponentially increasing number of parameters for its documentation.

Kohn's statement has led to focus electronic structure calculations for large systems on methods which do not require knowing the many-electron wavefunction. It has provided a strong argument for density functional theory (DFT) \cite{Hohenberg1964,Kohn1965} which does not attempt to calculate many-body wavefunctions. DFT has been highly successful and in fact has revolutionized electronic structure calculations in solid state physics. The price has been that it is not free of arbitrariness.

The above exponential wall problem (EWP), does not appear when the electron interactions are treated in mean-field approximation. In this case the $N$-electron Schr\"odinger equation reduces to a one-electron problem in a self-consistent potential \cite{Hartree1928}. 

The limitation of Schr\"odinger's wavefunction to $N \leq N_0$ is highly unsatisfactory. There are no indications that the wave character of an interacting electron system differs when $N \gtrless N_0$. By compromizing on the accuracy, local correlation methods were developed which scale linearly with electron number $N$. See, e.g., the PNO (pair-natural orbitals) coupled-cluster approach of Neese and coworkers \cite{Neese14}.

When searching for the origin of this limitation one realises that it is the multiplicative property of the wavefunction when two independent subsystems $A$ and $B$ are considered, i.e., in second quantisation
\begin{equation}
| \psi (A,B) \rangle =  | \psi (A) \rangle \otimes | \psi (B) \rangle ~~~.
\label{eq:02}
\end{equation}
For illustration consider a chain of $L$ sites with $N = 2L$ electrons. The ground state $| \psi_0 \rangle$ is of the general form
\begin{equation}
| \psi_0 \rangle =  \sum\limits_{i_1, \dots, i_L} C_{i_1, \dots, i_L} | i_1 \rangle \otimes | i_2 \rangle \otimes \dots \otimes | i_L \rangle 
\label{eq:03}
\end{equation}
where $| i_\nu \rangle$ denotes the different electronic configurations on site $\nu$. When we disentangle the sites more and more so that they become uncoupled for all practical purposes, the matrix in front factorizes, i.e.,
\begin{equation}
C_{i_1 \dots i_L} = c_{i_1} \cdot c_{i_2} \cdot ~\dots ~ \cdot c_{i_L} ~~~.
\label{eq:04}
\end{equation}
In this case consider the ground state with two electrons on each site. Assume that these two interacting electrons are described by $m$ configurations so that the total number of configurations is $m^L$ and thus increases exponentially with $L$ or $N$. Yet, the information contained in $| \psi_0 \rangle$ is merely given by the amplitudes of the configurations on a single site. In this limit the exponential growth of the configurations is accompanied by an increasing redundancy of information. 

The above example shows that a generalization of the concept of wavefunctions for $N > N_0$ must avoid configurations with redundant information. This is the case for disentangled subsystems $A$ and $B$ when we postulate that the total wavefunction is 
\begin{equation}
| \psi (A, B) \rangle = | \psi (A) \rangle \oplus | \psi (B) \rangle~~~,
\label{eq:05}
\end{equation}
i.e., when the wavefunctions of independent subsystems are added rather than multiplied. This is the case when instead of $\psi$ we consider $\ln \psi$. 

In order to see that this postulation relates to the action function we recall how Schr\"odinger derived his equation \cite{Schroedinger1926}. When searching for an equation describing matter waves by a field $\Phi$, he started from an ansatz
\begin{equation}
\Phi = e^{iW/\hbar} = e^{i(-Et+R)/\hbar}~~~,
\label{eq:06}
\end{equation}
where the action function $W=-Et+R$ contains the energy of the system and a time $t$ independent part $R$.

Schr\"odinger asked which form $R$ must have in order that $\Phi$ obeys a wave equation. He found that when $R$ is expressed in terms of the logarithm of another matter field $\psi$, i.e., 
\begin{equation}
R = -i \hbar \ln \psi~~~,
\label{eq:07}
\end{equation}
then $\Phi$ will obey a wave equation, provided that the field $\psi$ fulfills Schr\"odinger's equation. Note that the action $R$ is additive with respect to independent subsystems $A$ and $B$ while $\psi$ is multiplicative. Thus in order to eliminate the EWP we have to characterize wavefunctions via the action function $R$, i.e., via $\ln \psi$. These wavefunctions are additive with respect to inxependent subsystems.

When applied to the ground state of the chain (\ref{eq:03}) the number of configurations is only $mL$. There is no redundant information in this limit and the electron number is not limited to $N < N_0$. It turns out that calculations with the logarithm can be avoided by making use of cumulants. Kubo \cite{Kubo1962} has pointed out their usefulness long time ago.

The cumulant $c$ of a matrix element of products of operator $A_n$ to power $\alpha$ is defined through
\begin{equation}
\label{eq:08}
\left< \phi_1 \left| A_1 \dots A_n \right| \phi_2 \right>^c = \frac{\partial}{\partial \lambda_1} \dots \frac{\partial}{\partial \lambda_n} ln  \left< \phi_1 \left| \prod^n_{i = 1} e^{\lambda_i A_i} \right| \phi_2 \right>_{\lambda_i = \dots = \lambda_n = 0} 
\end{equation}
and it is assumed that $\langle \phi_1 | \phi_2 \rangle \neq 0$. The averaging vectors $| \phi_\nu \rangle$ are members of the $N$-electron Hilbert space. Equation (\ref{eq:08}) is independent of their norm. By setting $A_1 = \dots = A_n = A$, multiplying with $\lambda^n/n!$ and summing over $n$ we obtain from Eq. (\ref{eq:08})
\begin{equation}
\label{eq:09}
\ln \langle \phi_1 | e^{\lambda A} | \phi_2 \rangle = \langle \phi_1 | e^{\lambda A} -1 | \phi_2 \rangle^c~~~.
\end{equation}
Cumulants eliminate all factorizable contributions when matrix elements are evaluated by operator contractions. Note the similarity to Green's functions when disconnected diagrams are discarded \cite{Goldstone1957}. In the following we show how the ground-state wavefunction of a ''closed shell'' interacting electron system is modified when the logarithm of Schr\"odinger's wavefunction is used. We describe the system by a Hamiltonian $H$ which we decompose into a mean-field part $H_0$ and a remaining residual interaction part $H_1$, i.e. $H = H_0 + H_1$. The EWP is caused by $H_1$. We assume that the ground-state wavefunction of $H_0$ is known from Schr\"odinger's equation and denote it by $| \Phi_0 \rangle$, i.e., $H_0 | \Phi_0 \rangle = E_0 | \Phi_0 \rangle$. It is advantageous to call it the {\it vacuum state} because of the absence of any fluctuations caused by the residual interactions $H_1$. They generate vacuum fluctuations and cause the EWP.

The latter are particle-hole excitations represented by operators $A_i$. They generate new configurations the number of which increases exponentially with $N$. In order to obtain size extensive results for a given set of vacuum fluctuations the $A_i$ must enter the ground state in an exponential form \cite{Jastrow1955}, i.e.,
\begin{eqnarray}
\label{eq:10}
| \psi_0 \rangle & = & exp \left( \sum^M_{i=1} \tilde\lambda_i A_i \right) | \Phi_0 \rangle \\ \nonumber
& = & \tilde\Omega | \Phi_0 \rangle~~~.
\end{eqnarray}
This is a Coupled Cluster (CC) ansatz \cite{Coester1960,Bishop1978,Evangelista18} but with a subtile difference which we want to mention. In CC theory the holes which are created by particle-hole excitations are always in orbitals occupied in $| \Phi_0 \rangle$ while the created electron are in states unoccupied in $| \Phi_0 \rangle$. This restriction is not necessary in Eq. (\ref{eq:10}). Note that $| \psi_0 (\tilde\lambda_i) \rangle$ and $| \psi_0 (\tilde\lambda_i + \delta \lambda_i) \rangle$ are orthogonal when $N > N_0$ for all practical purposes because of the EWP. The prefactor $\tilde\Omega$ in Eq. (\ref{eq:10}) is the M\o ller operator. It transforms the mean-field ground state $| \Phi_0 \rangle$ into the ground state $| \psi_0 \rangle$ of $H$ provided $N<N_0$. For $N > N_0$ it looses its meaning. Often one writes $\tilde\Omega = 1 + \tilde S$ in which case $\tilde S$ acts like a scattering matrix.

For small electron numbers $N$ the operator $\tilde\Omega$ can be determined within a given set of basis functions by a full configuration interaction (full CI) calculations \cite{Goldstone1957}. For larger values of $N < N_0$ various wavefunction based quantum-chemical methods are available, e.g. M\o ller-Plesset perturbation expansions, Coupled Cluster, Coupled Electron Pair approximations (CEPA), Projection and Partitioning methods to name a few \cite{Loewdin1986,Kutzelnigg1975,Meyer1971,Ahlrichs1979,Fulde12}.        

The goal is to modify $\tilde\Omega$ so that the restriction $N < N_0$ becomes obsolete. We do this by taking the logarithm of $\tilde\Omega$ when defining $| \psi_0 \rangle$. Thus, from Eqs. (\ref{eq:09} - \ref{eq:10})
\begin{eqnarray}
\label{eq:11}
\left< \phi_j \left| \Omega \right| \Phi_0 \right> & = & \ln \left< \phi_j \left| \tilde\Omega \right| \Phi_0 \right> \\ \nonumber
& = & \ln \left< \phi_j \left| {\rm exp} \left( \sum_i \lambda_i A_\nu \right) \right| \Phi_0 \right> \\ \nonumber
& = & \left< \phi_j \left| {\rm exp} \left( \sum_i \lambda_i A_i \right) -1 \right| \Phi_0 \right>^c \\ \nonumber
& = & \left< \phi_j \left| {\rm exp} \sum_i \lambda_i A_i  \right| \Phi_0 \right>^c ~~~.
\end{eqnarray}
Here we have used that cumulants are independent of the norm of the averaging vectors so that $\langle \phi_j | 1 | \Phi_0 \rangle^c = \ln \langle \phi_j | \Phi_0 \rangle = 0$ when $\langle \phi_j | \Phi_0 \rangle = 1$ is chosen.

The resulting Eq. (\ref{eq:11}) is surprisingly simple. The only difference between $\Omega$ and $\tilde\Omega$ is that in the former case cumulants have to be used when matrix elements involving $\Omega$ are treated. This is enough to eliminate the EWP and to ensure Eq. (\ref{eq:05}). Until now we have not yet specified how the $\lambda_i$ are determined. We expect that the $\tilde\lambda_i$ and $\lambda_i$ will not coincide.

We recall briefly from \cite{Fulde12} how the ground-state wavefunction $| \psi_0 \rangle$ of $H$ is obtained from $\Omega$. We start from
\begin{eqnarray}
\label{eq:12}
\nabla_{\Phi_0} \left< \Phi_0 \left| 1 \right| \psi_0 \right>^c 
& = & \nabla_{\Phi_0} \ln \left< \Phi_0 \left| \psi_0 \right. \right> \\ \nonumber
& = & \frac{\left. \left| \psi_0 \right. \right>}{\left< \Phi_0 \left| \psi_0 \right. \right>} = \left. \left| \psi_0 \right.  \right>_{\rm norm}~~~.
\end{eqnarray}
By setting $| \psi_0 \rangle = \Omega | \Phi_0 \rangle$ we obtain
\begin{equation}
\label{eq:13}
\left. \left| \psi_0 \right.  \right>_{\rm norm} = \nabla_{\Phi^L_0} \left< \Phi_0 \left| \Omega \right| \Phi_0 \right>^c
\end{equation}
where the upper script $L$ implies that the gradient is with respect to the left $\Phi_0$.

By using the logarithm of the solution of Schr\"odinger's equation like in the corresponding action function we have found a simple way to eliminate the EWP. Wavefunctions of separated subsystem have become additive this way instead of multiplicative. The use of cumulants requires the following bilinear form in Liouville or operator space
\begin{equation}
\label{eq:14}
(A|B) = \left< \Phi_0 | A^+ B | \Phi_0 \right>^c~~~.
\end{equation}
With this redefinition of operator products the dimensions of Hilbert space do not grow exponentially with electron number $N$ and we do not have to worry about size extensivity. Therefore we may simply limit ourselves to a linearised form of the fluctuations. This implies the ansatz
\begin{eqnarray}
\label{eq:15}
| \Omega ) & = & \left. \left| 1 + \sum_i  \eta_i A_i \right. \right) \\ \nonumber
& = & \left. \left| 1 + S \right. \right)
\end{eqnarray}  
where $S$ is the cumulant scattering matrix. The parameters $\eta_i$ can be determined from the identities
\begin{equation}
\label{eq:16}
(A_i | H \Omega) = 0
\end{equation}
because the left hand side factorizes and therefore the cumulants vanish. For high quality results or when correlations are strong one must go, of course beyond a linear Ansatz Eq. (\ref{eq:15}).

At this stage we can make connection with Refs. \cite{Kladko1998,Fulde19} where we had derived Eq. (\ref{eq:15}) by a sequence of infinitesimal transformations in Hilbert space, taking us from $| \Phi_0 \rangle$ to $| \psi_0 \rangle$. The present derivation, which relates the many-body character of the wavefunction to the action function provides a simpler physical picture for the removal of the EWP, and is the main result of the present communication.

We have now a method at hand to calculate electronic structures of large systems with controlled approximations, in particular for periodic solids. This was questioned before, see e.g. \cite{Kohn1999}.

It is well known that an attractive feature of the cumulant scattering operator $S$ is that it can be decomposed into different parts, because it is additive. A possible choice is
\begin{equation}
\label{eq:17}
| S \rangle = \sum_I | S_I ) + \sum_{\langle IJ \rangle} | S_{I,J} - S_I - S_J) + \dots
\end{equation}
where the indices $I, J, ...$ refer to different sites (or bonds) \cite{Fulde12,Kladko1998,Fulde19,Paulus06}. When any of these terms in calculated, e.g., the increment $| S_{IJ})$, all electrons in $| \Phi_0 \rangle$ are kept frozen, except those on sites $I$ and $J$. The latter are correlated by applying any of the quantum chemical methods used when $N < N_0$. 

There exist a number of studies how the cumulant scattering matrix $|S)$ is evaluated in practice by starting from the decomposition Eq. (\ref{eq:17}). Usually the calculations start from a set $L$ of basis functions (e.g., Gaussian-type orbitals). The size of $L$ affects, of course, the possible accuracy of the calculated correlation contributions of the cumulant scattering matrix $|S)$ \cite{Fulde12,Fulde19,Paulus06}.

An advantage of the decomposition (\ref{eq:17}) is that one can study separately the correlations of different large systems by limiting the sites $I, J, \dots$ to the spatial parts of the system one is interested in.

We conclude with the hope that the introduction of the action function is providing a simple physical picture for Eq. (\ref{eq:17}). It was derived before in a more abstract way (see, e.g. \cite{Fulde19}).

\vspace{1cm}

\section*{Acknowledgements}

I would like to thank Hermann Stoll and Peter Thalmeier for numerous helpful discussions.

\end{document}